\documentclass[useAMS,usenatbib]{mn2e}
\usepackage[dvips]{graphicx}
\usepackage{url}
\usepackage{amsmath}

\title[Broad-line type Ic Supernova SN 2014ad]{Broad-line type Ic Supernova SN 2014ad}

\author[D. K. Sahu et al.]
{D.K. Sahu$^1$,
G.C. Anupama$^1$,
N.K. Chakradhari$^2$,
S. Srivastav$^1$,
Masaomi Tanaka$^3$,
\newauthor 
Keiichi  Meada$^4$,
Ken'ichi Nomoto$^{5,6}$\\
1. Indian Institute of Astrophysics, II Block Koramangala, Bangalore 560034, India\\
2. Pt. Ravishankar Shukla University, Raipur 492010, India\\
3. National Astronomical Observatory Japan, Mitaka, Tokyo 181-8588, Japan\\
4. Department of Astronomy, Kyoto University, Kitashirakawa-Oiwake-cho, Sakyo-ku, Kyoto 606-8502, Japan\\
5. Kavli Instt. for the Physics and Mathematics of the Universe, (WPI),  The University of Tokyo, Kashiwa, Chiba 277-8583, Japan\\
6. Hamamatsu Professor\\
E-mail : dks@iiap.res.in (DKS), gca@iiap.res.in (GCA), nkchakradhari@gmail.com (NKC), ssrivastav@iiap.res.in (SS),\\ masaomi.tanaka@nao.ac.jp (MT), keiichi.maeda@kusastro.kyoto-u.ac.jp (KM), nomoto@astron.s.u-tokyo.ac.jp (KN)}

\begin{document}

\date{Accepted .....; Received ......}

\pagerange{\pageref{firstpage}--\pageref{lastpage}} \pubyear{2014}

\maketitle

\label{firstpage}

\begin{abstract}
We present  optical and ultraviolet  photometry, and low resolution optical spectroscopy of the broad-line
type Ic supernova SN 2014ad in the galaxy PGC 37625 (Mrk 1309), covering 
the evolution of the supernova during $-$5 to +87 d with respect to the date of maximum in $B$-band. 
A late phase spectrum obtained at +340 d is also presented. With an absolute $V$ band magnitude 
at peak of $M_{V}$ = $-$18.86 $\pm$ 0.23 mag, SN 2014ad is fainter
than Gamma Ray Burst (GRB) associated supernovae, and brighter than most of the normal 
and broad-line type Ic supernovae without an associated GRB. The spectral evolution indicates the expansion velocity of the ejecta, as measured using the Si\,{\sc ii}  line, to be as high as $\sim$ 33500 km\,s$^{-1}$  around maximum, while during the post-maximum phase it settles down at  $\sim$ 15000 km\,s$^{-1}$. 
The expansion velocity of SN 2014ad is  higher than all other well observed broad-line type Ic supernovae except the GRB associated SN 2010bh. The explosion parameters, determined by applying the Arnett's analytical light curve model  to the observed bolometric light curve, indicate that it was an energetic explosion with a kinetic energy of $\sim$ (1 $\pm$ 0.3)$\times$10$^{52}$ ergs,  a total ejected mass of $\sim$ (3.3 $\pm$ 0.8) M$_\odot$, and $\sim$ 0.24 M$_\odot$ of $^{56}$Ni was synthesized in the explosion.  The metallicity of the host galaxy near the supernova region is estimated to be $\sim$ 0.5 Z$_\odot$. 
\end{abstract}  


\section{Introduction}
\label{sec04ab_intro}
Core-collapse supernovae (CCSNe) arising from progenitors which are stripped of hydrogen and/or 
helium are known as  stripped-envelope  supernovae (SE-SNe). This includes  
types IIb, Ib and Ic supernovae (SNe).  For type IIb SNe, the progenitor retains a thin layer of hydrogen at 
the time of explosion, whereas in type Ib, the hydrogen envelope is completely removed. Type Ic 
SNe show neither hydrogen nor helium in their spectra  around maximum brightness, indicating 
that  both hydrogen and helium envelopes of the progenitor stars are removed before the explosion
(\citealt{fili97,tura03a}). The exact nature of the  progenitors and the process operational in removing the 
outer envelopes is not fully  understood. The progenitors of type Ic SNe are thought to be either a 
massive Wolf-Rayet (WR) star, or a less massive  star in a binary system. In the case of WR stars, the helium envelope is removed by powerful stellar winds, and in a binary system, mass transfer to the companion helps in removing the helium envelope (see \citealt{lang12} for a recent review). 

A relatively  small fraction ($\sim$ 4\% ;  \citealt{shiv17}) of  type Ic SNe  show very broad lines in their spectra obtained close to maximum light, indicating very high expansion velocity ($\sim$ 15000--30000 km\,s$^{-1}$) of the ejecta.  
These are known as broad-line type Ic SNe.   The association of the nearby GRB 980425 \citep{gala98} with SN 1998bw indicated that the GRB associated SNe are of broad-line Ic type. This was later established by subsequent  discovery of many other broad-line SNe associated with GRBs/XRFs (\citealt{hjor03,stan03,male04,pian06,bufa12,toy16}). 
 The broad-line type Ic SNe associated with GRB/XRF, are also known as  engine driven supernovae. These have a rapidly rotating central compact object powered by accretion, and often associated with relativistic outflow.  However, there are  some broad-line SNe which are not associated with GRBs/XRFs (\citealt{sand12, mazz13,walk14}).  A possible interpretation of the absence of an observed GRB with some broad-line type Ic SNe is a relativistic jet initially beamed away from   the line of sight of the observer \citep{rhoa99}.   As  the  decelerating jet spreads laterally, the emission becomes effectively isotropic and shifts to  longer wavelengths. Irrespective of the initial viewing angle, the  afterglow emission in radio is expected to increase rapidly on a time scale of few weeks to several years (\citealt{pern98,waxm04}). Thus late phase observation in radio can be used to search for an evidence of off-axis GRB (\citealt{sode06}, and references therein). The non-detection of late time radio emission from  a sample of type Ibc SNe, including broad-line type Ic   led \citet{sode06} to conclude that every broad-line type Ic supernova does not harbour a GRB.    

In a systematic study by \citet{modj16}, it was shown that type Ic SNe with GRBs have higher expansion velocities as well as broader line widths than SNe Ic without an observed GRB. Based on this, it was suggested that  the broad-line SNe Ic without an observed GRB may have had lower energy and/or choked jet that imparted lower velocities to the supernova ejecta.  Recently, two energetic broad-line Ic SNe 2009bb \citep{pign11} and 2012ap \citep{mili15}, showed mildly relativistic ejecta coupled with strong radio emission. The pre-maximum spectra of both the objects show the presence of helium, leading \citet{marg14} to suggest the jet might have failed because it was damped by the additional helium layer in the progenitor. These objects appear to act as a bridge between the highly relativistic, collimated GRBs and more normal type Ic SNe.    

SN 2014ad was discovered on March 12.4, 2014 by the Catalina Real-Time Transient Survey, in PGC 37625 (Mrk 1309) at RA = 11$^{\text h}$ 57$^{\text m}$ 44$^{\text s}$.44, Dec = $-$10$^{\circ}$ 10$'$ 15$''$.7  \citep{howe14}. The object was located at  3.5$''$ West and 7.2$''$ North  of the center of the host galaxy. \citet{howe14} reported  that the early  spectrum taken on March 14.9 and 15.9 with the South African Large Telescope, showed a blue continuum and broad lines with minima near 4400 \AA\  and 5100 \AA.  It was suggested that it could be peculiar type Ia supernova, either a type Iax event or similar to SN 2002bj and SN 2005ek, or perhaps something unique \citep{howe14}.  Further spectroscopic observations  of this object on March 16.9, 17.9 and 18.9 revealed the development of  broad lines and its resemblance to broad-line type Ic SNe like SN 1998bw and  SN 2002ap,  though with some differences in the line velocities and shapes. Recently \citet{stev17} have presented spectropolarimetric study of SN 2014ad. In this paper we present results of optical imaging and spectrosopy, together with the {\it Swift} UVOT data from the archives.

\begin{figure}
\centering   
\resizebox{\hsize}{!}{\includegraphics{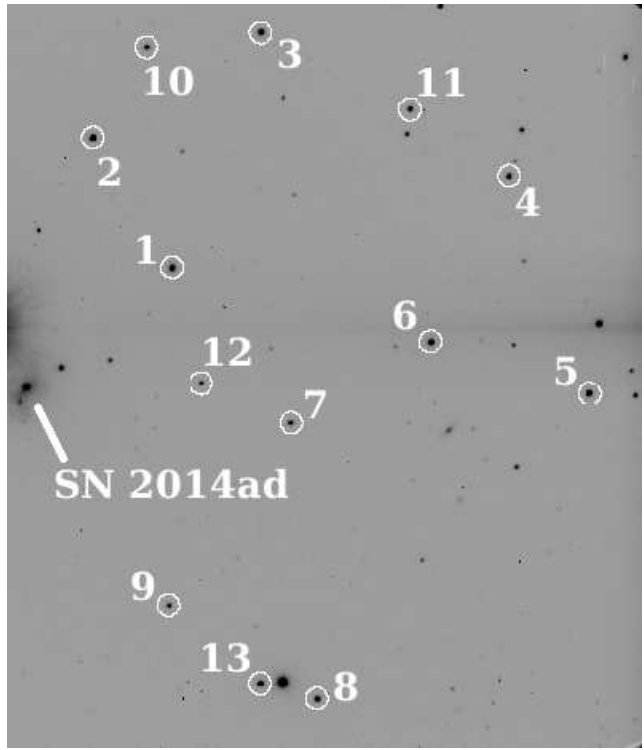}} 
\caption[]{$R$ band image of SN 2014ad in the galaxy Mrk 1309. North is up and east to the left. The field 
stars marked with numbers 1--13 are used as local standards. The supernova is shifted close to the  edge to avoid a nearby very bright star.}
\label{fig_field}
\end{figure}

\section{The light curve}
\label{sec:light}
\subsection{Optical observations}

Optical imaging of SN 2014ad was carried out  in Bessell's $U$, $B$, $V$, $R$ and $I$ bands, using the Himalayan Faint Object Spectrograph Camera (HFOSC) mounted on the Himalayan Chandra Telescope (HCT). The photometric monitoring of this object started on  2014 March 13 (JD 245,6730.41), soon after the discovery and was continued until 2014 June 11 (JD 245,6820), when the object went into Solar conjunction. The central 2K$\times$2K region of 2K$\times$4K pixels CCD chip was used for imaging observations. With a plate scale of 0.296 arcsec\,pixel$^{-1}$, the central 2K$\times$2K pixels cover a field of view  of 10$\times$10 arcmin$^2$. Gain and readout noise of the CCD camera are 1.22 electron\,ADU$^{-1}$ and 4.87 electrons, respectively. Further details about the telescope and instrument can be obtained from  \url{http://www.iiap.res.in/centers/iao}.  On each night several calibration frames {\it e.g.} bias frames, twilight sky flat frames  were taken.  The observed  data  was reduced in a standard manner,  using various tasks available within Image Reduction and Analysis Facility ({\sc iraf}{\footnote {{\sc iraf} is distributed by the National Optical Astronomy Observatories, which are operated by the Association of Universities for Research in Astronomy, Inc., under cooperative agreement with the National Science Foundation}}) package. 

Photometric standard regions PG1323-085,  PG1525-071, PG1633+099 and PG2213-006 from the list of \citet{land92} were observed on 2014 March 21, March 29, April 27, May 17 and June 04 under  photometric sky conditions. 
Aperture photometry  was performed to estimate the instrumental magnitudes of  Landolt's standard stars. The aperture photometry was done at two apertures, at an optimal aperture determined using the  aperture growth curve and an aperture  close to full width half maximum (FWHM) of stellar profile. Bright stars in the field were used to determine  an aperture correction between the two different apertures.  This correction was applied to the magnitude obtained at smaller aperture.  The average extinction  co-efficients for the site \citep{stal08}, were used for atmospheric  extinction  correction.  The average colour terms for the system  were used  to determine the photometric zero points  by fitting a linear relation between the observed  and standard colours. A sequence of secondary standards in the supernova field,  was calibrated using 
the estimated zero points and average colour terms.  Calibrated $UBVRI$ magnitudes of the secondary standards, averaged over 5 nights are given in Table \ref{tab_std} and have been marked in Fig. \ref{fig_field}.  To avoid saturation due to a very bright star next to the host galaxy, the supernova was placed close to the  edge of the field of view. 
 
The aperture and profile fitting photometry of the supernova suffer due to the complex background of the host galaxy at supernova location.  Subtraction of  the host galaxy image (template subtraction) works well  in removing the galaxy background and results in  accurate photometry.  Deep template images of the host galaxy were obtained with the same instrumental setup, once the supernova faded sufficiently. 
The template images in different bands were  subtracted from the corresponding supernova frames and 
aperture photometry was performed on the residual frames to estimate instrumental magnitude 
of the supernova.  The magnitudes of the  secondary standards were measured using aperture  photometry. Finally,  the calibrated magnitudes of the supernova were obtained differentially  with respect to the  set of secondary standard stars in the supernova field. The supernova magnitudes in $U$, $B$, $V$, $R$ and $I$  bands have been listed in Table \ref{sn_mag}. The errors reported in Table \ref{sn_mag} are estimated by taking into account the photometric  error (computed
by {\sc iraf}) and the calibration error.

\begin{table*}
\caption{Magnitudes of secondary standard stars in the field of SN 2014ad. The stars are marked in Fig. \ref{fig_field}.}
\centering
\begin{tabular}{@{}lccccc@{}}
\hline
ID & U  & B & V &  R & I \\
\hline
1  & 15.559 $\pm$ 0.023&   15.160 $\pm$ 0.015&   14.370 $\pm$ 0.017&    13.895 $\pm$ 0.022&   13.464 $\pm$ 0.016\\
2  & 15.194 $\pm$ 0.016&   15.009 $\pm$ 0.020&   14.356 $\pm$ 0.004&    13.963 $\pm$ 0.016&   13.607 $\pm$ 0.009\\
3  & 15.717 $\pm$ 0.021&   15.544 $\pm$ 0.022&   14.863 $\pm$ 0.008&    14.458 $\pm$ 0.029&   14.072 $\pm$ 0.025\\
4  & 16.078 $\pm$ 0.016&   15.759 $\pm$ 0.018&   15.021 $\pm$ 0.014&    14.590 $\pm$ 0.014&   14.196 $\pm$ 0.017\\
5  & 15.271 $\pm$ 0.007&   15.387 $\pm$ 0.017&   14.838 $\pm$ 0.005&    14.463 $\pm$ 0.006&   14.076 $\pm$ 0.018\\
6  & 16.319 $\pm$ 0.031&   15.891 $\pm$ 0.014&   15.116 $\pm$ 0.013&    14.672 $\pm$ 0.018&   14.264 $\pm$ 0.015\\
7  & 16.918 $\pm$ 0.016&   16.438 $\pm$ 0.017&   15.534 $\pm$ 0.015&    15.019 $\pm$ 0.022&   14.524 $\pm$ 0.017\\
8  & 16.008 $\pm$ 0.033&   15.928 $\pm$ 0.012&   15.298 $\pm$ 0.012&    14.920 $\pm$ 0.009&   14.542 $\pm$ 0.012\\
9  & 19.799 $\pm$ 0.099&   18.573 $\pm$ 0.003&   17.051 $\pm$ 0.024&    16.003 $\pm$ 0.011&   14.850 $\pm$ 0.013\\
10 & 16.746 $\pm$ 0.020&   16.228 $\pm$ 0.027&   15.421 $\pm$ 0.005&    14.925 $\pm$ 0.004&   14.481 $\pm$ 0.023\\
11 & 16.651 $\pm$ 0.009&   16.438 $\pm$ 0.022&   15.749 $\pm$ 0.006&    15.338 $\pm$ 0.018&   14.954 $\pm$ 0.022\\
12 & 16.739 $\pm$ 0.034&   16.829 $\pm$ 0.015&   16.290 $\pm$ 0.030&    15.944 $\pm$ 0.034&   15.611 $\pm$ 0.034\\
13 & 14.951 $\pm$ 0.033&   15.145 $\pm$ 0.002&   14.839 $\pm$ 0.005&    14.630 $\pm$ 0.012&   14.379 $\pm$ 0.012\\
\hline
\end{tabular}
\label{tab_std}
\end{table*}

\begin{table*}
\caption{Optical $UBVRI$ photometric observations of SN 2014ad with HCT.}
\centering
\begin{tabular}{@{}lcrccccc@{}}
\hline
 Date & JD$^a$ & Phase$^b$ & U & B & V & R & I\\
\hline
13/03/2014& 730.407&$-$5& 14.489 $\pm$ 0.010 &   15.237 $\pm$ 0.027&  15.099 $\pm$ 0.013 & 14.934 $\pm$ 0.025& 14.859 $\pm$ 0.028\\
14/03/2014& 731.423&$-$4&                  &   15.065 $\pm$ 0.030&  14.781 $\pm$ 0.015 & 14.697 $\pm$ 0.023& 14.706 $\pm$ 0.027\\
18/03/2014& 735.273&$0$ & 14.595 $\pm$ 0.020 &   14.705 $\pm$ 0.020&  14.084 $\pm$ 0.061 & 13.992 $\pm$ 0.043& 13.882 $\pm$ 0.082\\
19/03/2014& 736.206&+1 &                   &   14.756 $\pm$ 0.013&  14.105 $\pm$ 0.020 & 13.908 $\pm$ 0.020& 13.803 $\pm$ 0.019\\
21/03/2014& 738.376&+3 &  14.985 $\pm$ 0.018 &   15.000 $\pm$ 0.009&  14.003 $\pm$ 0.005 & 13.815 $\pm$ 0.019& 13.733 $\pm$ 0.004\\
25/03/2014& 742.316&+7 &  15.479 $\pm$ 0.015 &   15.257 $\pm$ 0.008&  13.93  $\pm$ 0.006 & 13.771 $\pm$ 0.010& 13.69  $\pm$ 0.006\\
26/03/2014& 743.317&+8 &  15.572 $\pm$ 0.061 &   15.392 $\pm$ 0.010&  13.960 $\pm$ 0.013 & 13.777 $\pm$ 0.006& 13.716 $\pm$ 0.008\\
27/03/2014& 744.250&+9 &  15.705 $\pm$ 0.029 &   15.669 $\pm$ 0.009&  14.074 $\pm$ 0.012 & 13.815 $\pm$ 0.007& 13.747 $\pm$ 0.013\\
28/03/2014& 745.217&+10&  15.919 $\pm$ 0.014 &   15.593 $\pm$ 0.006&  14.064 $\pm$ 0.007 & 13.726 $\pm$ 0.014& 13.769 $\pm$ 0.019\\
29/03/2014& 746.212&+11&  15.973 $\pm$ 0.019 &   15.784 $\pm$ 0.005&  14.142 $\pm$ 0.011 & 13.866 $\pm$ 0.021& 13.787 $\pm$ 0.011\\
30/03/2014& 747.144&+12&  16.133 $\pm$ 0.018 &   15.803 $\pm$ 0.014&  14.202 $\pm$ 0.013 & 13.931 $\pm$ 0.009& 13.832 $\pm$ 0.014\\
01/04/2014& 749.385&+14&  99.999 $\pm$ 0.999 &   15.989 $\pm$ 0.010&  14.359 $\pm$ 0.011 & 14.048 $\pm$ 0.010& 13.923 $\pm$ 0.012\\
03/04/2014& 751.157&+16&  16.515 $\pm$ 0.007 &   16.165 $\pm$ 0.006&  14.489 $\pm$ 0.016 & 99.999 $\pm$ 0.999& 14.024 $\pm$ 0.013\\
09/04/2014& 757.151&+22&  16.925 $\pm$ 0.007 &   16.628 $\pm$ 0.008&  15.010 $\pm$ 0.010 & 14.593 $\pm$ 0.018& 14.369 $\pm$ 0.009\\
11/04/2014& 759.195&+24&                   &   16.616 $\pm$ 0.029&  15.074 $\pm$ 0.060 & 14.710 $\pm$ 0.010& 14.490 $\pm$ 0.013\\
14/04/2014& 762.179&+27&                   &   16.748 $\pm$ 0.042&  15.222 $\pm$ 0.019 & 14.901 $\pm$ 0.017& 14.627 $\pm$ 0.016\\
17/04/2014& 765.141&+30&  17.039 $\pm$ 0.017 &   16.902 $\pm$ 0.019&  15.407 $\pm$ 0.013 & 15.027 $\pm$ 0.018& 14.771 $\pm$ 0.017\\
18/04/2014& 766.094&+31&                   &   16.944 $\pm$ 0.010&  15.580 $\pm$ 0.017 & 15.107 $\pm$ 0.020& 14.819 $\pm$ 0.027\\
19/04/2014& 767.128&+32&                   &   16.943 $\pm$ 0.012&  15.481 $\pm$ 0.015 & 15.139 $\pm$ 0.022& 14.843 $\pm$ 0.025\\
23/04/2014& 771.104&+36&  17.259 $\pm$ 0.018 &   17.019 $\pm$ 0.019&  15.717 $\pm$ 0.019 & 15.308 $\pm$ 0.019& 14.978 $\pm$ 0.022\\
27/04/2014& 775.162&+40&  17.523 $\pm$ 0.016 &   17.334 $\pm$ 0.027&  15.918 $\pm$ 0.015 & 15.481 $\pm$ 0.022& 15.106 $\pm$ 0.024\\
30/04/2014& 778.113&+43&  17.500 $\pm$ 0.014 &   17.199 $\pm$ 0.018&  15.900 $\pm$ 0.018 & 15.509 $\pm$ 0.019& 15.160 $\pm$ 0.021\\
03/05/2014& 781.162&+46&  17.611 $\pm$ 0.014 &   17.423 $\pm$ 0.014&  15.936 $\pm$ 0.010 & 15.598 $\pm$ 0.010& 15.226 $\pm$ 0.013\\
07/05/2014& 785.108&+50&  17.583 $\pm$ 0.023 &   17.340 $\pm$ 0.012&  16.036 $\pm$ 0.031 & 15.685 $\pm$ 0.023& 15.298 $\pm$ 0.025\\
14/05/2014& 792.112&+57&  17.784 $\pm$ 0.035 &   17.419 $\pm$ 0.016&  16.241 $\pm$ 0.007 & 15.870 $\pm$ 0.012& 15.489 $\pm$ 0.022\\
15/05/2014& 793.135&+58&  17.766 $\pm$ 0.054 &   17.389 $\pm$ 0.018&  16.232 $\pm$ 0.010 & 15.894 $\pm$ 0.021& 15.508 $\pm$ 0.012\\
17/05/2014& 795.110&+60&  17.718 $\pm$ 0.018 &   17.372 $\pm$ 0.010&  16.323 $\pm$ 0.012 & 15.906 $\pm$ 0.013& 15.538 $\pm$ 0.012\\
04/06/2014& 813.130&+78&  17.989 $\pm$ 0.038 &   17.540 $\pm$ 0.013&  16.627 $\pm$ 0.010 & 16.300 $\pm$ 0.014& 15.949 $\pm$ 0.006\\
11/06/2014& 820.162&+85&  17.937 $\pm$ 0.080 &   17.657 $\pm$ 0.026&  16.861 $\pm$ 0.011 & 16.469 $\pm$ 0.021& 16.005 $\pm$ 0.034\\
\hline      
\multicolumn{8}{@{}l}{$^a$245 6000+; $^b$Observed phase with respect to the epoch of $B$ band maximum: JD = 245,6735.11.}
\end{tabular}		            
\label{sn_mag}	       
\end{table*}

\subsection{{\it Swift} UVOT observations} 

The ground based data on SN 2014ad is supplemented by data obtained by the Ultra Violet Optical Telescope (UVOT) on-board the {\it Swift} satellite, and retrieved from the {\it Swift} data archive.
SN 2014ad was  observed during  JD 245,6735 to JD 245,6754 in optical broad band filter {\it u} (3465 \AA) and three UV  filters {\it uvw2} (1928 \AA), {\it uvm2} (2246 \AA) and {\it uvw1} (2600 \AA). The data were reduced using various modules in HEASoft (the High Energy Astrophysics Software) following \citet{pool08} and \citet{brow09}. The magnitude of the supernova was obtained using {\it uvotsource} task. This task performs aperture photometry at a user defined aperture after taking into account the coincidence losses. Aperture photometry with an aperture of 5 arcsec is recommended. However,  as the supernova was faint, photometry at a smaller aperture of 3 arcsec,  having high signal-to-noise ratio, is performed and  aperture correction provided  by \citet{pool08} is applied to the extracted magnitudes. The background  counts are estimated from the nearby region, using an aperture of size similar to that used for the supernova. {\it Swift} photometry of the supernova is reported in Table \ref{snmag_uvot}.  

\begin{table*}
\caption{UV/Optical photometric observations of SN 2014ad with {\it Swift} UVOT.}
\centering
\begin{tabular}{@{}lcrcccc@{}}
\hline
 Date & JD$^a$ & Phase$^b$ & $uvw2$ & $uvm2$ & $uvw1$ & $u$ \\
\hline
19/03/2014& 735.69&0& 17.667 $\pm$ 0.19& 17.346 $\pm$ 0.13&  16.010 $\pm$ 0.08& 14.730 $\pm$ 0.04 \\
21/03/2014& 738.10&3& 17.159 $\pm$ 0.14& 17.154 $\pm$ 0.11&  16.127 $\pm$ 0.08& 14.971 $\pm$ 0.04 \\
23/03/2014& 740.36&5& 17.233 $\pm$ 0.14& 17.223 $\pm$ 0.08&  16.467 $\pm$ 0.09& 15.244 $\pm$ 0.04 \\
25/03/2014& 741.93&7& 17.428 $\pm$ 0.18& 17.515 $\pm$ 0.11&  16.560 $\pm$ 0.10& 15.651 $\pm$ 0.06 \\
27/03/2014& 743.96&9&                  & 16.653 $\pm$ 0.10&  15.872 $\pm$ 0.16&                   \\
29/03/2014& 746.02&11&17.349 $\pm$ 0.16& 17.608 $\pm$ 0.15& 17.076 $\pm$ 0.14& 16.074 $\pm$ 0.06 \\
31/03/2014& 747.91&13&17.183 $\pm$ 0.14& 17.566 $\pm$ 0.17& 17.062 $\pm$ 0.18& 16.090 $\pm$ 0.11 \\
03/04/2014& 751.29&16&17.617 $\pm$ 0.17& 17.591 $\pm$ 0.15& 17.331 $\pm$ 0.20& 16.293 $\pm$ 0.11 \\
04/04/2014& 751.63&16&18.037 $\pm$ 0.25& 17.503 $\pm$ 0.13& 17.182 $\pm$ 0.16&                \\
06/04/2014& 754.04&19&17.496 $\pm$ 0.13& 17.369 $\pm$ 0.11& 17.059 $\pm$ 0.12& 16.529 $\pm$ 0.10 \\
\hline
\multicolumn{7}{@{}l}{$^a$245,6000+; $^b$Observed phase with respect to the epoch of $B$ band maximum: JD = 245,6735.11.}
\end{tabular}			    
\label{snmag_uvot}	    
\end{table*}

\subsection{Light curve evolution}
\label{sec:light_evol}

\begin{figure}
\centering
\resizebox{\hsize}{!}{\includegraphics{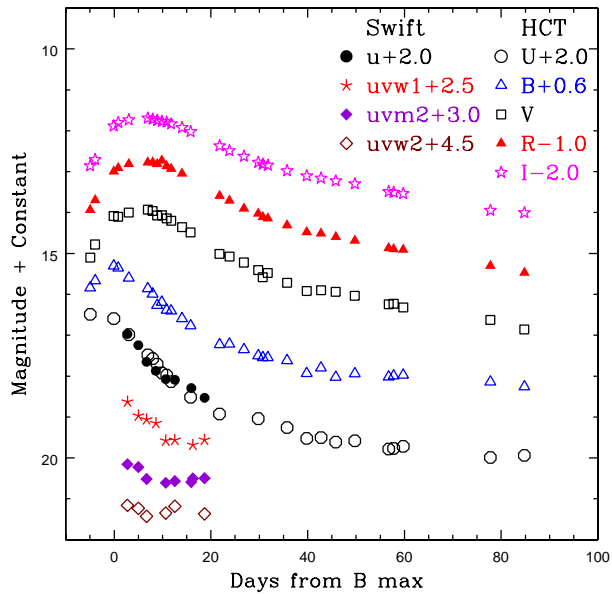}}
\caption[]{Optical $UBVRI$ and {\it Swift} light curves of SN 2014ad. The light 
curves have been shifted by the amount indicated in the legend. Typical error
in  photometry is less than the point size.}
\label{fig_lc}
\end{figure}

The optical and UV light curves of SN 2014ad are  presented  in Fig. \ref{fig_lc}.
The optical observations started well before the phase of maximum light in different
bands,  while the {\it Swift} UV observation covers only the first 20 days since $B$ maximum. 
The light curve evolution of SN 2014ad in the $B$ and $U$ bands is similar and shows a fast decline soon after peak, whereas the $V$, $R$ and $I$ light curves exhibit a slow decline.   The $uvm2$  and $uvw2$ light curves do not show any considerable evolution and they are almost flat. A similar flat behaviour was reported in the $uvw1$ band light curve  of SN 2009bb \citep{pign11}, and was interpreted  as due to the dominance of the background flux a few days after $B$ maximum. In the absence of reference images in the {\it Swift} UV filters, a proper background subtraction is difficult. 

The date of maximum light in different bands  is estimated by fitting a cubic spline to the observed data points around maximum light. The peak in  bluer bands occurred earlier than the redder bands. 
The  date of maximum and peak magnitude in the $U$ band could not be 
determined,  as we do not have enough pre-maximum data points.
The peak in the $B$ band occurred on JD 245,6735.11$\pm$0.25 with an  apparent magnitude 
14.787 $\pm$ 0.025 mag. The peak in the $V$ band occurred at $\sim$ 5.5 days  and the peak in the $R$ and $I$ bands occurred at  $\sim$ 6.2 days  after the $B$ band maximum. 

SN 2014ad was discovered on March 12.40, 2014 (JD 245,6728.9) at an apparent magnitude 
of 15.7 mag. It was  not detected  in  the image of the  host galaxy obtained  on 
March 06.37 (JD 245,6722.87),   down to a limiting magnitude of 19.0 mag. This can be used to 
constrain the  date of explosion of SN 2014ad  as JD 245,6725 $\pm$ 3 days. Further, the 
observed dates of maximum  brightness indicate a rapid rise to maximum in the  $B$ band 
in 10 $\pm$ 3 days and the $R$ band rise time as 16 $\pm$ 3 days.

\begin{figure}
\centering
\resizebox{\hsize}{!}{\includegraphics{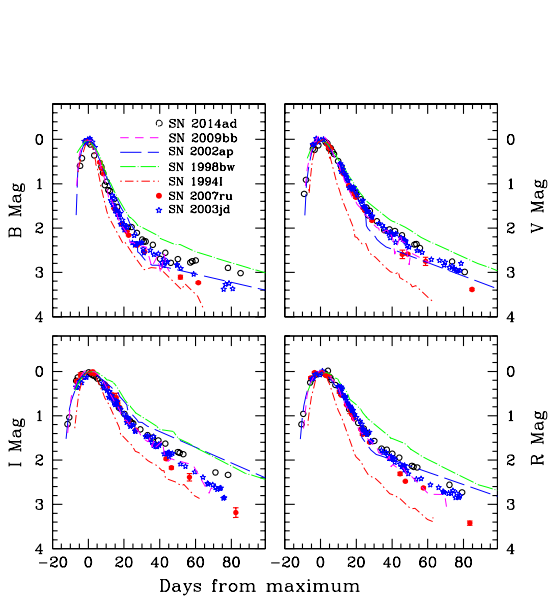}}
\caption[]{Light curve comparison of SN 2014ad with other well studied broad-line and normal Ic SNe.}
\label{fig_lccomp}
\end{figure}

The light curve of SN 2014ad is plotted along with other well studied  SNe Ic SN 1994I \citep{rich96};  broad-line SNe Ic SN 2009bb \citep{pign11}, SN 2007ru \citep{sahu09}, SN 2003jd \citep{vale08}, SN 2002ap (\citealt{fole03, pand03}) and GRB 980425/SN 1998bw  \citep{gala98} in Fig. \ref{fig_lccomp}. The rising part of the light curve of
SN 2014ad is faster than SN 1998bw and similar to those of SN 2002ap and SN 2009bb. The light curve decline in 15 days from the peak is estimated as $\Delta m_{15}(B)$ = 1.31$\pm$0.03, $\Delta m_{15}(V)$ = 0.95$\pm$0.06, $\Delta m_{15}(R)$ = 0.77$\pm$0.03 and $\Delta m_{15}(I)$ = 0.62$\pm$0.02. 

The light curve of SN 2014ad in all the bands is broader than type Ic SN 1994I. 
Except for SN 1998bw, the post-maximum evolution of the light curve of SN 2014ad is 
slower than other SNe used for comparison. During the early post-maximum phase ($<$ 20 days after maximum in the $B$-band) $B$, $V$ and $R$ band light curves of SN 2014ad match well with SN 1998bw and after that  SN 2014ad declines faster than SN 1998bw. The $I$ band light curve of SN 2014ad is similar to that of SN 2007ru and  SN 2003jd, and  narrower than SN 1998bw. The $V$ band decline rate $\Delta m_{15}(V)$   of SN 2014ad is comparable to that of  SN 2002ap, faster than SN 1998bw  and slower than SN 2003jd,  SN 2007ru and SN 2009bb.

\begin{figure}
\centering
\resizebox{\hsize}{!}{\includegraphics{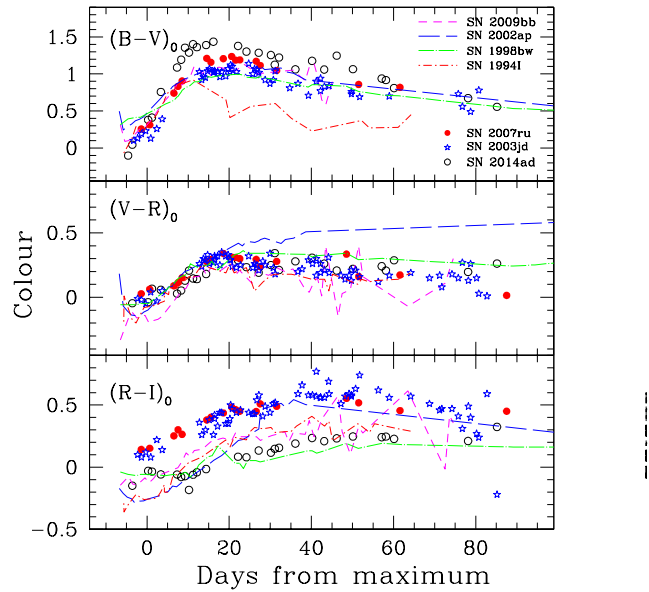}}
\caption[]{Colour curve evolution of SN 2014ad compared with other well studied broad-line and normal Ic SNe.}
\label{fig_colcomp}
\end{figure}

The de-reddened colour curves of SN 2014ad are plotted
in Fig. \ref{fig_colcomp} with those of SN 2009bb, SN 2007ru, SN 2003jd,
SN 2002ap, SN 1998bw and SN 1994I. The colour curves
of SN 2014ad have been corrected for colour excess $E(B-V)$ = 0.22 mag
 (refer Section \ref{sec:reddening}) while $E(B-V)$ value for other supernovae
is taken from their respective references. The  $(B-V)$ colour  evolution of SN 2014ad is similar to other broad-line type Ic SNe in comparison and different from the  type Ic supernova SN 1994I. The $(V-R)$ colour evolution of SN 2014ad is similar to other broad-line Ic and  type Ic SNe. 
The $(R-I)$ colour of type Ic SNe   shows considerable  dispersion, with SN 2014ad lying at the bluer end.
 The  colour evolution of SN 2014ad is similar to that  of SN 1998bw.

\subsection{Reddening, distance and absolute magnitudes}
\label{sec:reddening}
The Galactic reddening in the direction of Mrk 1309,  estimated from the infrared dust map of \citet{schl98} is  $E(B-V)$ = 0.045 mag. The low resolution spectrum  of SN 2014ad around epoch of maximum
light shows a strong Na\,{\sc i}\,D absorption line due to interstellar matter within the Milky Way, 
 and a weaker absorption at the rest wavelength of the host galaxy. The five spectra of SN 2014ad,  obtained
close to maximum light, were averaged to improve the signal-to-noise ratio for measuring the
equivalent width of the Na\,{\sc i}\,D absorption line.
The average equivalent width of the Galactic 
component  of the Na\,{\sc i}\,D line is 0.98 $\pm$ 0.06 \AA\ and from the host galaxy is 0.06 $\pm$ 0.01 \AA. 
Using the relation between the equivalent width of Na\,{\sc i}\,D line and reddening \citep{poz12}, 
we estimate  $E(B-V)_\text{Gal}$ =  0.20 $\pm$ 0.06 mag,    $E(B-V)_\text{host}$ = 0.02 mag and total reddening  
$E(B-V)_\text{total}$ =  0.22 $\pm$ 0.06 mag. The estimated  $E(B-V)_\text{Gal}$ is high as compared to the value 
reported by \citet{schl98}. The estimated higher  $E(B-V)_\text{total}$ of SN 2014ad is  
consistent with  colour evolution, the reddening corrected $(B-V)$ and $(V-R)$ colours of
SN 2014ad are similar to other broad-line type Ic SNe (see Fig. \ref{fig_colcomp}).  
For further analysis, we have adopted   $E(B-V)_\text{total}$ =  0.22 $\pm$ 0.06 mag. 

The distance to the host galaxy Mrk 1309 of SN 2014ad is estimated using the radial velocity
of the host,  corrected for the local group infall onto the Virgo cluster
V$_\text{Virgo}$ = 1904 $\pm$ 32 km\,s$^{-1}$ \citep{moul00}.  
For $H_{0}$ = 72 km\,s$^{-1}$\,Mpc$^{-1}$, the distance to SN 2014ad is 26.44 Mpc 
and distance modulus is 32.11 $\pm$ 0.15 mag.

 The peak absolute magnitude of SN 2014ad in different bands has been estimated using  
the reddening and distance modulus mentioned above. With $V$ band
absolute magnitude of $-$18.86 $\pm$ 0.23 mag, SN 2014ad is more luminous than  type 
Ic SNe 1994I  ($M_{V}$ = $-$17.62 $\pm$ 0.3 mag; \citealt{rich96, saue06}), 
2004aw ($M_{V}$ = $-$18.02 $\pm$ 0.3 mag; \citealt{taub06}), 
2007gr ($M_{V}$ = $-$17.22 $\pm$ 0.18 mag; \citealt{hunt09}), 
broad-line  SNe 2012ap ($M_{V}$ = $-$18.67 $\pm$ 0.08 mag;  \citealt{mili15}),  
2009bb ($M_{V}$ = $-$18.65 $\pm$ 0.34 mag; \citealt{pign11}), 
2002ap ($M_{V}$ = $-$17.37 $\pm$ 0.05 mag; \citealt{fole03, pand03}), and
XRF 060218/SN 2006aj ($M_{V}$ = $-$18.67 $\pm$ 0.08 mag; \citealt{modj06}) and 
fainter than GRB 980425/SN 1998bw ($M_{V}$ = $-$19.12 $\pm$ 0.05 mag; \citealt{gala98}),
GRB 031203/SN 2003lw ($M_{V}$ = $-$19.75 $\pm$ 0.5 mag; \citealt{male04}). The  
$V$ band absolute magnitude of SN 2014ad is, however, comparable to those of broad-line SNe 
2007ru  ($M_{V}$ = $-$19.06 $\pm$ 0.2 mag; \citealt{sahu09}) and 
2003jd ($M_{V}$ = $-$18.9 $\pm$ 0.3  mag; \citealt{vale08}).
Our estimate of $M_{V}$ for SN 2014ad is consistent with that reported by \citet{stev17}.

The $R$ band absolute magnitude of SN 2014ad  $-$18.87 $\pm$ 0.3 mag, is  comparable 
to those of broad-line Ic SNe (M$_{R}$ = $-$19.0 $\pm$ 1.1 mag) and engine 
driven SNe (M$_{R}$ = $-$18.9 $\pm$ 0.4 mag), and  brighter than normal Ic 
SNe  (M$_{R}$ = $-$18.3 $\pm$ 0.6 mag) from the  sample of \citet{drou11}.

\section{The optical spectrum}
\label{sec:spec}
\subsection{Observations}
\label{sec:spec_observation}

Spectroscopic observations of SN 2014ad were performed during  2014 March 13 (JD 245,6730.35)  to  
2014 June 13 (JD 245,6822.17) using the HCT.    To cover the entire  optical waveband,  the observations were carried out using grisms Gr\#7  (wavelength  range 3500--7800 \AA) and Gr\#8 (5200--9250 \AA), available with the HFOSC. The spectral resolution is  $\sim$ 7 \AA. A late phase spectrum was obtained with the 8.2 m 
Subaru Telescope equipped with the Faint Object Camera and Spectrograph (FOCAS; \citealt{kash02}) on 2015 February 22. A 0$''$.8 slit and B300 grism, covering 4700--9000 \AA, were used for this observation.   The journal of spectroscopic observations is given in Table  \ref{tab:spec_log}. Spectroscopic data reduction was carried out using tasks available within {\sc IRAF}.  The spectra were bias subtracted and  flat-fielded. The one
dimensional spectra were extracted using the optimal extraction method. The one dimensional spectra obtained with the HCT were wavelength calibrated  using the arc lamp spectra of FeAr and FeNe. The wavelength calibration for Subaru data was obtained  using  ThAr arc lamp spectrum. The wavelength calibrated spectra were cross checked using bright night sky emission lines and whenever required, a small shift was applied.  The spectrum of spectrophotometric standard stars,  observed on the same night, was used  to correct for the instrumental response and flux calibrate the supernova spectra. 
The spectra in the two different regions were combined, scaled to a weighted mean, to give the  final spectrum on a relative flux scale.  Except for the late phase spectrum obtained using Subaru, other spectra  were  brought to an absolute flux scale using the $UBVRI$ magnitudes.  The supernova spectra were corrected for the host galaxy redshift of $z$ = 0.01546 (from NED) and dereddened by the total reddening $E(B-V)_\text{total}$ = 0.22 mag as
estimated in Section \ref{sec:reddening}. The telluric lines have not been removed from the spectra.
\begin{table}
\centering
\caption{Log of spectroscopic observations of SN 2014ad.}
\begin{tabular}{@{}lccc@{}}
\hline
Date & JD$^a$ & Phase$^b$& Range (\AA) \\
\hline
13/03/2014 & 730.35& $-$5&  3500-7800; 5200-9250\\
14/03/2014 & 731.34& $-$4&  3500-7800; 5200-9250\\
18/03/2014 & 735.31&    0&  3500-7800; 5200-9250\\
19/03/2014 & 736.22&   +1&  3500-7800; 5200-9250\\
21/03/2014 & 738.40&   +3&  3500-7800; 5200-9250\\
25/03/2014 & 742.34&   +7&  3500-7800; 5200-9250\\
27/03/2014 & 744.22&   +9&  3500-7800; 5200-9250\\
28/03/2014 & 745.37&  +10&  3500-7800; 5200-9250\\
29/03/2014 & 746.34&  +11&  3500-7800; 5200-9250\\
31/03/2014 & 748.12&  +13&  3500-7800; 5200-9250\\
04/04/2014 & 752.20&  +17&  3500-7800; 5200-9250\\
14/04/2014 & 762.13&  +27&  3500-7800; 5200-9250\\
16/04/2014 & 764.25&  +29&  3500-7800; 5200-9250\\
23/04/2014 & 771.30&  +36&  3500-7800; 5200-9250\\
27/04/2014 & 775.23&  +40&  3500-7800; 5200-9250\\
03/05/2014 & 781.14&  +46&  3500-7800; 5200-9250\\ 
17/05/2014 & 795.13&  +60&  3500-7800; 5200-9250\\ 
27/05/2014 & 805.19&  +70&  3500-7800; 5200-9250\\
04/06/2014 & 813.14&  +78&  3500-7800; 5200-9250\\ 
13/06/2014 & 822.17&  +87&  3500-7800; 5200-9250\\ 
22/02/2015$^c$ &1075.43 & +340   & 4700-9000\\
\hline
\multicolumn{4}{@{}l@{}}{$^a$245,6000+; $^b$in days relative to $B$ band maximum.}\\
\multicolumn{4}{@{}l@{}}{$^c$Observation taken with Subaru Telescope.}
\end{tabular}
\label{tab:spec_log}
\end{table}

\subsection{Spectral Evolution}
\label{sec:spec_evol}
\subsubsection{Pre-maximum spectral evolution}
\begin{figure}
\centering
\resizebox{\hsize}{!}{\includegraphics{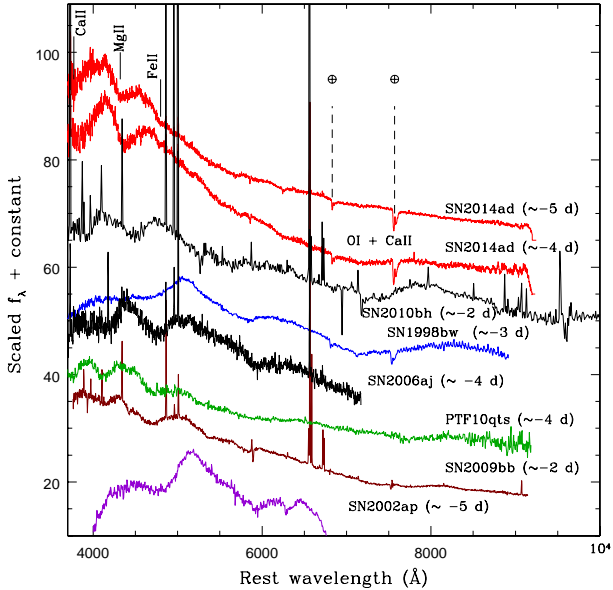}}
\caption[]{Pre-maximum spectra of SN 2014ad, compared with those of other broad-line type
Ic SNe at similar epoch. For clarity the spectra have been shifted vertically. Telluric lines are marked with a circled plus sign.}

\label{fig_spec1}
\end{figure}

The spectral evolution of SN 2014ad in the pre-maximum phase is 
presented in Fig \ref{fig_spec1}. 
The first two spectra obtained on $-$5 and $-$4 d with respect to the $B$ band maximum show
a blue continuum with two distinct broad absorptions at $\sim$ 3800 \AA\  and  at $\sim$ 4400 \AA. 
Weak notches are also seen at $\sim$ 4800 \AA\  and at $\sim$ 5800 \AA\ in both the spectra.  
The absorption at  $\sim$ 3800 \AA\ is due to Ca\,{\sc ii},  and the absorptions at $\sim$ 4400 \AA\ and $\sim$ 4800\AA\ are due to Mg\,{\sc ii} and Fe\,{\sc ii}, respectively \citep{walk14}.  In the  $-$4 d spectrum, a depression is seen at  $\sim$ 7300 \AA\, which was not present in the spectrum taken on day $-$5. This feature is due to heavily blended  O\,{\sc i} and Ca\,{\sc ii} NIR triplet, and requires presence of sufficient material at velocity  higher than  30000 km\,s$^{-1}$  (Mazzali et al. 2002). The apparent broad emission-like features  at 4000 \AA\ and 4600 \AA\  in the early phase spectra of SN 2014ad  do not result from discrete emission lines, but they are merely the regions of low opacity, from where photons have
higher probability of escaping \citep{iwam98,mazz00}.  

The pre-maximum  spectra of SN 2014ad is compared with the  spectrum of other broad-line type Ic 
SNe 1998bw \citep{pata01}, PTF10qts \citep{walk14}, 2009bb \citep{pign11}, 2010bh \citep{bufa12} and 2002ap \citep{fole03} obtained at similar epoch,  in Fig. \ref{fig_spec1}.  
The shape of the spectrum of SN 2014ad is very similar to the spectra of PTF10qts, SN 2010bh and SN 2009bb, however, the features in the spectrum of SN 2014ad are blueshifted more.   
Except  SN 2002ap, all other SNe  show a blue continuum in the pre-maximum phase. There are 
noticable differences between the spectrum of SN 2014ad and SN 1998bw:  the spectrum of SN 1998bw 
does not show the pseudo-emission peak at $\sim$ 4400 \AA\ seen in  SN 2014ad, which is due to higher metal content at high velocity \citep{mazz13}. The 
pseudo-emission peak at $\sim$ 6300 \AA\ seen in the  spectrum of SN 1998bw is absent in the spectrum of SN 2014ad and the strong  absorption at $\sim$ 7300 \AA\ is  less pronounced in the spectrum of SN 2014ad. Features similar to  those seen in SN 2014ad have also been noted in the pre-maximum spectra of GRB 130702A/SN 2013dx \citep{deli15}. 

\subsubsection{Spectral  evolution during maximum and  post-maximum phase} 
\begin{figure}
\centering
\resizebox{\hsize}{!}{\includegraphics{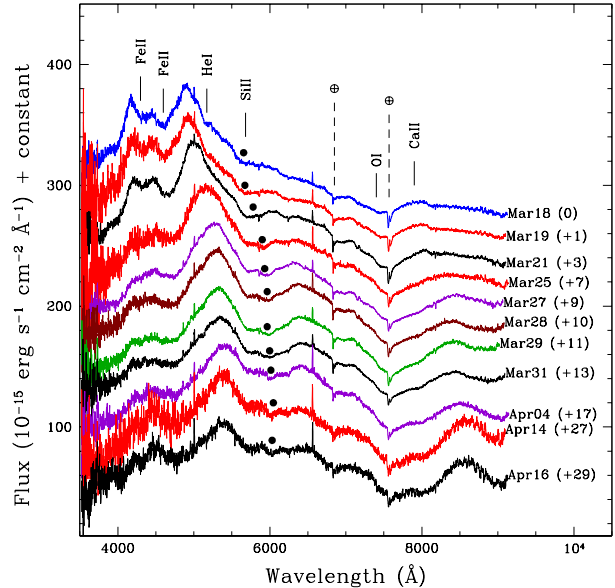}}
\caption[]{Spectral evolution  of SN 2014ad during early post-maximum phase. The black dot shows the  position of Si\,{\sc ii} 6355\AA\ line.
Telluric lines are marked with a circled plus sign.}
\label{fig_spec2}
\end{figure}

The spectral evolution during the first month, when the  light curve shows steep decline from maximum light, is shown in Fig. \ref{fig_spec2}.  The spectrum at $B$ band maximum shows  a significant evolution  as compared to the pre-maximum spectra. The 
 continuum becomes redder with emission peaks  at  $\sim$ 4200 \AA\ and $\sim$ 4900 \AA.  
  The peak at  $\sim$ 4200 \AA\ weakens after first 
few days, whereas the peak  at $\sim$ 4900 \AA\ continues to be there with a gradual redshift. 
A small number of broad features are seen in the spectrum, which become increasingly dominant.
 
The main features in the spectra are due to lines of Si\,{\sc ii}, O\,{\sc i}, Ca\,{\sc ii} and Fe\,{\sc ii}.
The weak notch seen in the pre-maximum spectra at $\sim$ 5800 \AA\ appears to become stronger, which
is most likely due to Si\,{\sc ii}  6355 \AA\  line. In the redder part of the spectrum, the O\,{\sc i} line becomes stronger and  is heavily blended with the Ca\,{\sc ii} NIR triplet. 

The spectrum of SN 2014ad obtained close to maximum light shows a  depression  at $\sim$ 5200 \AA. A similar feature was noticed in the spectrum of SN 2009bb, and   was  identified with He\,{\sc i}  5876 \AA\ line \citep{pign11}. If this feature is indeed due to He\,{\sc i} 5876 \AA\ feature, it corresponds to a velocity of 37000 km\,s$^{-1}$, which is consistent with 
 the photospheric  velocity deduced from the absorption trough of Si\,{\sc ii} 6355 \AA\ line (refer Section \ref{sec:phot_vel}). The broad absorption feature between 7300 \AA\ to 8000 \AA\ is similar to that seen in SN 2012ap, identified with multiple components of Ca\,{\sc ii}, including a detached component of high velocity (HV) and very high velocity (VHV) \citep{mili15}.

\begin{figure}
\centering
\resizebox{\hsize}{!}{\includegraphics{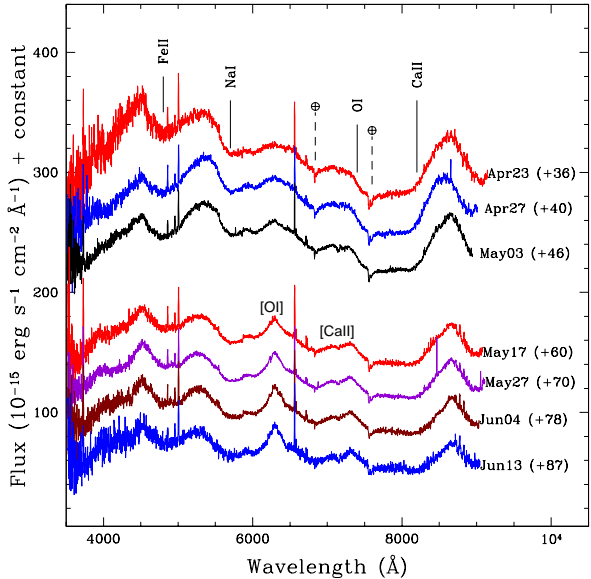}}
\caption[]{Spectral evolution  of SN 2014ad during post-maximum phase. Telluric lines are marked with a circled plus sign.} 
\label{fig_spec3}
\end{figure}

\begin{figure}
\centering
\resizebox{\hsize}{!}{\includegraphics{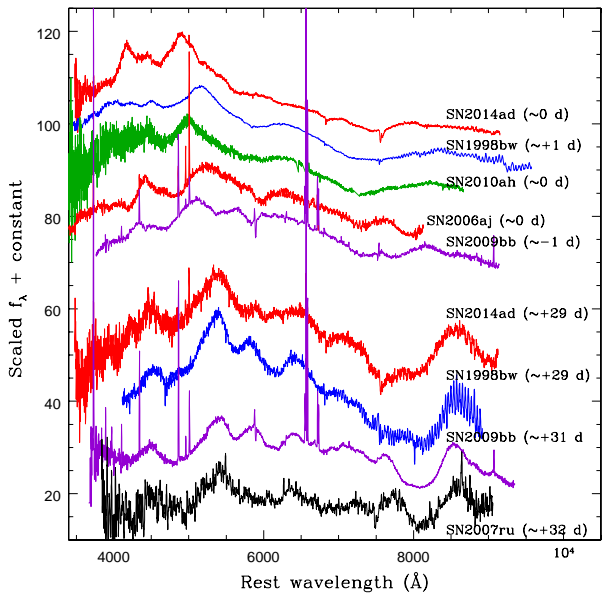}}
\caption[]{Spectrum  of SN 2014ad at $B$ band maximum  and $\sim$ one month after maximum, 
compared with those of other broad-line type
Ic SNe at similar epoch. For clarity the spectra have been shifted vertically.}
\label{fig_spec_comp2}
\end{figure}

Spectral evolution of SN 2014ad  during +36 d to +87 d, is shown in Fig. \ref{fig_spec3}. 
During this phase the absorption dominated spectrum gradually changes to an emission dominated spectrum, indicating the ongoing transition to  the nebular phase.  The blue part of the spectrum gets increasingly supressed and the strength of the emission component  of Ca\,{\sc ii} NIR triplet increases. The absorption due to Na\,{\sc i} D line becomes increasingly stronger.  All the lines are still broad, indicating persistence of high expansion velocity of the ejecta.  

The  characteristic nebular lines due to [Mg\,{\sc i}]  4571 \AA, [O\,{\sc i}]  6300, 6363 \AA\  and [Ca\,{\sc ii}]  7291, 7324 \AA, superimposed on the continuum are  seen in our   spectrum  obtained on +87 d.  The presence of continuum shows that the transition to the nebular phase is not yet complete.  

The spectrum of SN 2014ad around maximum light   is compared  with the spectrum of other broad-line objects obtained at similar epoch in Fig. \ref{fig_spec_comp2}. 
With broad spectral features, the spectrum of SN 2014ad is very similar to those of SN 1998bw and SN 2010ah.
 A comparison of  +30 d spectrum of SN 2014ad  with those of SN 1998bw, SN 2007ru and  SN 2009bb, shows that    the spectra of all the SNe in comparison look similar except for the difference in the line broadening due to differences in the expansion velocities (refer Fig. \ref{fig_spec_comp2}). They are all dominated by the P-Cygni profile of Ca\,{\sc ii} NIR triplet. The O\,{\sc i} 7774 \AA\ line in the spectra of SN 2014ad and SN 1998bw is weak, it is even weaker in SN 2007ru, but in SN 2009bb it is stronger as compared to SN 2014ad.

\subsubsection{Spectral  evolution during nebular  phase} 
\begin{figure}
\centering
\resizebox{\hsize}{!}{\includegraphics{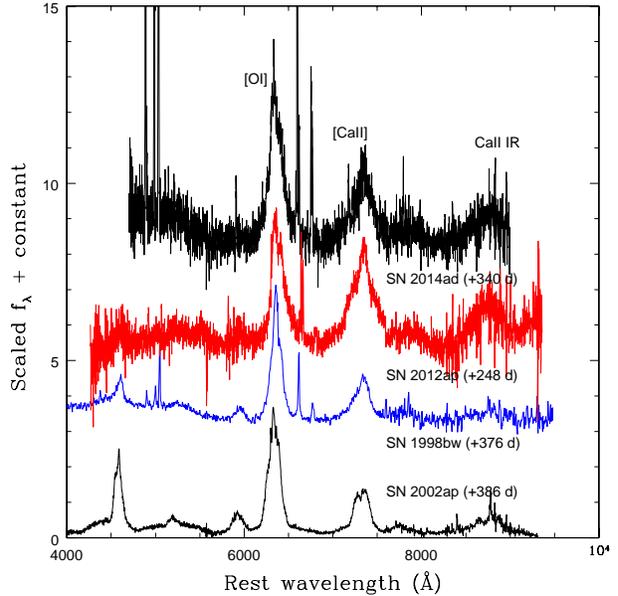}}
\caption[]{Nebular spectrum of SN 2014ad plotted with that of SN 1998bw, SN2002ap and SN 2012ap. 
 For clarity the spectra have been shifted vertically.}
\label{fig_spec_comp6}
\end{figure}

 The  spectrum of SN 2014ad obtained at +340 d is plotted in Fig. \ref{fig_spec_comp6}. This spectrum is dominated by [O\,{\sc i}]  6300, 6363 \AA\ and [Ca\,{\sc ii}]  7291, 7324 \AA\, possibly blended with [O\,{\sc ii}] 7320, 7330 \AA\ forbidden emission lines. Narrow emission lines from the host galaxy are also prominently seen in this spectrum. Nebular spectrum of SN 1998bw, SN 2002ap and SN 2012ap,  around  similar epoch, have also been plotted in the same figure for comparison. 

The    [O\,{\sc i}]  6300, 6363 \AA  /[Ca\,{\sc ii}]  7291, 7324 \AA\   emission line ratio  is found to be 1.54. This line ratio is  important as it can be used as a good diagnostic of the main-sequence mass (M$_\text{MS}$) of the progenitor star \citep{maed07}. The mass of O in the ejecta of core-collapse SNe is very sensitive to the M$_\text{MS}$ of the progenitor and it increases with M$_\text{MS}$, whereas mass of the explosively synthesized Ca is insensitive to the M$_\text{MS}$ of the progenitor, making the  [O\,{\sc i}]/[Ca\,{\sc ii}]  line ratio higher  for massive progenitor star \citep{nomo06}.
 \citet{kunc15} have compiled the [O\,{\sc i}]/[Ca\,{\sc ii}]  line ratio for several core-collapse SNe of similar age during nebular phase. The line ratio is $<$ 1 for type IIP SNe included in their plot.   For stripped envelope core-collapse SNe, \citet{kunc15} find that   depending on whether the progenitor star is in a binary system or
a single star, the ratio is found to be different. The [O\,{\sc i}]/[Ca\,{\sc ii}]  line ratio for SN 2014ad (1.54) is comparable to those of SN 1998bw (1.7), SN 2002ap (2)  SN 2007ru (1.6) and higher  than SN 2009bb (0.8)  and SN 2012ap (0.9).  A   higher  value of [O\,{\sc i}]/[Ca\,{\sc ii}]  for SN 2014ad indicates towards higher M$_\text{MS}$ for the progenitor star.

\subsection{Photospheric velocity}
\label{sec:phot_vel}
In type Ib/c SNe the expansion velocity estimtated using the Fe\,{\sc ii} lines around 5000 \AA, is treated as representative of the photospheric velocity, as the other lines are formed much above the photosphere. 
The high expansion velocity of the ejecta in broad-line Ic SNe results in  blending  of Fe\,{\sc ii} lines making its identification difficult. In such cases,  the expansion velocity measured  using Si\,{\sc ii} line is generally used as photospheric velocity.

\begin{figure}
\centering
\resizebox{\hsize}{!}{\includegraphics{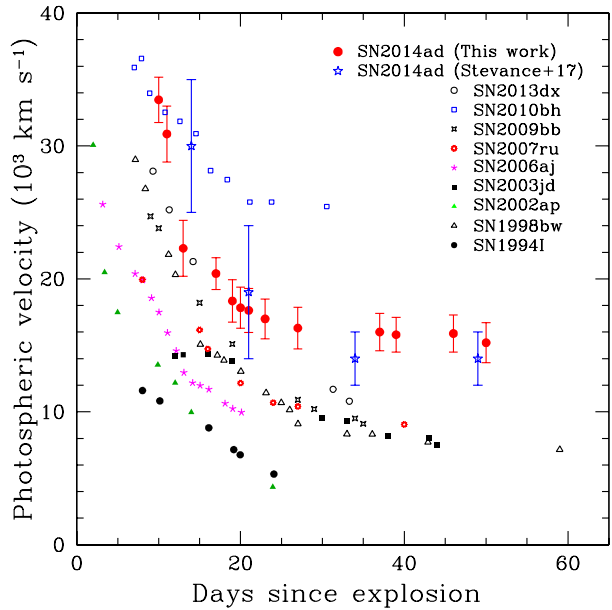}}
\caption[]{Temporal evolution of photospheric velocity  of SN 2014ad.  The photospheric velocity measured 
by \citet{stev17} is also shown. For comparison the Si\,{\sc ii} line velocities of other type Ic SNe are  plotted.}
\label{phot_vel}
\end{figure}

The photospheric velocity of SN 2014ad is  measured  
using absorption trough of the Si\,{\sc ii}  6355 \AA\ line, in the redshift corrected spectra.  During the pre-maximum phase,  due to severe blending of the spectral lines,  identification of even  Si\,{\sc ii} feature is difficult. Hence, velocity during the pre-maximum phase could not be measured. The position of Si\,{\sc ii} feature is marked in the spectra shown in Fig. \ref{fig_spec2}, and  temporal evolution  of  photospheric velocity is plotted in Fig. \ref{phot_vel}.  The  photospheric velocity of SN 2014ad, estimated  by fitting the Fe\,{\sc ii} blend using {\sc syn++}  \citep{stev17} is also plotted in  Fig. \ref{phot_vel}. Our  photospheric velocity  is consistent with estimates of    \citet{stev17}. 
The Si\,{\sc ii} line  velocity ranges from $\sim$ 33500 km\,s$^{-1}$ at $B$ maximum to $\sim$ 15000 km\,s$^{-1}$ at around 40 d after $B$ maximum.   The Si\,{\sc ii} line velocity of SN 2014ad showed steep decline  from $\sim$ 33500 km\,s$^{-1}$  to $\sim$ 21000 km\,s$^{-1}$ during $\sim$ 10 to 15 d after the explosion.  A similar drop  in the expansion velocity during the first 20 days after explosion has already been noticed in other broad-line type Ic SNe.

The  photospheric velocity of other supernovae have also been plotted in Figure   \ref{phot_vel} for comparison.
  We find that the velocity  of ejecta in SN 2014ad is unusually high.  Except for SN 2010bh,  expansion velocity of SN 2014ad is higher than all supernovae used in comparison.  
Beyond $\sim$ 20 days after explosion, the expansion velocity of SN 2014ad flattens at $\sim$ 15000 km\,s$^{-1}$, while it flattens at $\sim$ 25000 km\,s$^{-1}$ for SN 2010bh, and at $\sim$ 8000 km\,s$^{-1}$ for the other SNe.

\subsection{Spectral fitting with \sc syn++ }

\begin{figure}
\centering
\resizebox{\hsize}{!}{\includegraphics{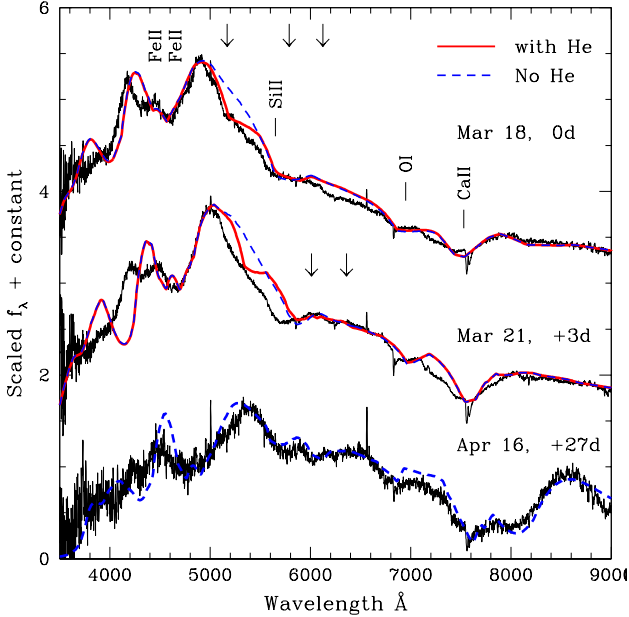}}
\caption[]{Spectra of SN 2014ad at  0 d, +3 d and +27 d,  plotted along with the synthetic fits generated using SYN++. The inclusion of He\,{\sc i} improves the fit at $\sim$ 5400 \AA. The expected position of  He\,{\sc i} lines 5876 \AA, 6678 \AA \ and 7065 \AA \  are shown  with  arrow marks.}
\label{synfits}
\end{figure}

We used  the {\sc syn++} code, an enhanced version of parameterized supernova spectrum synthesis code {\sc synow} \citep{fish00},   to generate synthetic spectra for SN 2014ad.   The  observed spectrum of SN 2014ad at  maximum light is fit well with the synthetic spectrum having a photospheric velocity of 35000 km\,s$^{-1}$ and black body temperature  of 13000 K (See Fig. \ref{synfits}). The synthetic spectrum includes ions of O\,{\sc i},  Mg\,{\sc ii}, Si\,{\sc ii}, Ca\,{\sc ii}, Fe\,{\sc ii} and Co\,{\sc ii}.  An excitation temperature of 6000 K was used for all the ions. Desired strength of the lines were obtained by varying the optical depth, which is an exponential  function of velocity. It is noticed that the synthetic spectrum with the above species, does not fit the observed spectrum properly in the region 5000--5600 \AA \  (shown by dashed blue line), especially the weak absorption around 5200 \AA.  This absorption minimum can be  reproduced by introducing   He\,{\sc i},  marginally detached from the photosphere (shown by solid red line),  at a velocity of $\sim$ 40000 km\,s$^{-1}$.  The expected locations of other He\,{\sc i} lines 6678 \AA\ and  7065 \AA\ are also  marked in the spectrum. However, as these lines are usually fainter as compared to  5876 \AA\  line, it is difficult to detect them.  The observed flux in the spectrum of SN  2014ad in the region bluer than 4000 \AA\ reduces drastically. Such a flux deficit in the synthetic spectrum could not be produced by Co\,{\sc ii}/Fe\,{\sc ii} ions but  adding Fe\,{\sc i} improves the fit.  

The best fit to the spectrum of SN 2014ad at +3 d is obtained at photospheric  velocity of 25000 km\,s$^{-1}$  and black body temperature of 10000 K.   The ions included  in the synthetic spectrum are the same as those used to fit the maximum light  spectrum with the same excitation temperature.  
Similar to the case of maximum light spectrum, the fit was examined with   and without adding He\,{\sc i}.  The  synthetic spectrum without   He\,{\sc i} appears to be matching well the observed spectrum.   This is consistent with the finding of \citet{stev17}, in which they did not find any indication  for  helium in the spectrum obtained at +4 d.  An additional high velocity component  of Ca\,{\sc ii} was included at 50000 km\,s$^{-1}$ to produce the broad feature of Ca\,{\sc ii}  NIR triplet.

The spectrum of SN 2014ad at +27 d  matches well with the synthetic spectrum having a  photospheric velocity of 16000 km\,s$^{-1}$ and black body temperature of 7000 K. The ions of O\,{\sc i}, Na\,{\sc i}, Si\,{\sc ii},  Ca\,{\sc ii}, Fe\,{\sc i}, Fe\,{\sc ii} and Co\,{\sc ii} are included at an excitation temperature of  6000 K each.   Broad and deep absorption feature of Ca\,{\sc ii} NIR triplet is reproduced by adding an additional high velocity component at 36000 km\,s$^{-1}$.  

The  presence of He\,{\sc i} line in the spectra of type Ic event SN 1994I was explored by \citet{fili95} and \citet{cloc96}, and its  possible contamination  with other species {\it e.g.} Na\,{\sc i}, C\,{\sc i} was discussed by \citet{mill99} and \citet{saue06}.  Detection of  He\,{\sc i}  has been reported in the broad-line type Ic SNe 2009bb \citep{pign11} and in 2012ap \citep{mili15}. The presence of He\,{\sc i} in SN 2009bb is inferred through  comparison of the optical spectra with the synthetic spectra,  inclusion of helium in the synthetic spectrum is found to fit the observed spectrum better. In SN 2012ap the detection of He\,{\sc i} in the optical spectra is supported by the presence of  10830 \AA\ and  20581 \AA\ lines.  The spectrum of SN 2010bh showed weak absorption features in the optical and NIR spectrum, compatible with He\,{\sc i}   5876 \AA\ and  10830 \AA\ lines.  However, the presence of  He\,{\sc i} in SN 2010bh could not be confirmed since the  20581 \AA\ line could not be clearly identified \citep{bufa12}. We have shown that inclusion of He\,{\sc i} in the synthetic spectrum close to $B$ band maximum  improves the fit marginally.  An unambiguous detection of helium  can be made only with NIR  spectra and detailed spectral modelling.   

\subsection{Metallicity of the host galaxy in the supernova region}

To understand the properties of the underlying region of the host galaxy, a spectrum of the supernova region was obtained on 2015 February 02, 
with the same settings as used for supernova observations.  Narrow emission  lines due to hydrogen, oxygen and sulphur from the host galaxy are clearly seen in the spectrum. The emission line fluxes were used to estimate the metallicity in the supernova region using different metallicity diagnostics available in the literature such as the calibration by \citet{mcga91}, and the $N2$ and $O3N2$ prescription of \citet{pett04}. The McGaugh calibration, which uses the $R_{23}$  ratio gives an oxygen abundance of 12 + $\log$ (O/H) = 8.38. The oxygen abundance estimated using the $N2$ and $O3N2$ index of \citet{pett04} is 8.42 and 8.37, respectively. Oxygen abundance was also estimated using the host galaxy emission lines in the nebular spectrum of  SN 2014ad, obtained  with the Subaru telescope.  It is found that it agrees well with the other estimates.   The average value of 12 + $\log$ (O/H) = 8.4$\pm$0.2, corresponds to  $\sim$ 0.5 Z$\odot$. This  shows that the metallicity at the supernova region is sub-solar. 
 \citet{modj08} have shown that GRBs are found in metal poor environment as compared to the broad-line type Ic SNe without GRBs.  It is also shown that,  following the strong line diagnostics  of \citet{kewl02},  the oxygen abundance 12 + $\log$ (O/H) at the location of the broad-line  SNe Ic  is $\sim$ 8.5, which corresponds to 0.2--0.6 Z$\odot$, depending on the adopted metallicity scale and solar abundance value.  Our estimates of metallicity at the location of SN 2014ad is consistent with findings of \citet{stev17} and  \citet{modj08}.

\section{Bolometric Light curve}
\label{sec:bol_curve}
The quasi-bolometric light curve of SN 2014ad was constructed using the optical and 
{\it Swift} UV magnitudes reported in Section 3. The observed magnitudes were corrected for
total extinction of $E(B-V)$ = 0.22 mag, using the Galactic reddening law of
\citet{card89}. The extinction corrected magnitudes were then  converted to 
monochromatic flux at the filter effective wavelength, using the magnitude to flux conversion zero 
points from  \citet{bess98}. The fluxes were then interpolated using the spline function and 
integrated from 0.31 $\micron$ to 0.93 $\micron$.  The quasi-bolometric luminosity thus obtained is
used to estimate the bolometric magnitude and is plotted in Fig. {\ref{bol_lc1_ad}. During  JD 245,6738  to JD 245,6751, when the  {\it Swift} UV  data was available, the integration was carried out from 0.16 $\micron$ to 0.93 $\micron$.  The bolometric light curve of  SN 2014ad  peaked on JD 245,6738 with bolometric magnitude 
-18.10 mag. The quasi-bolometric light curve of some broad-line type Ic SNe with and without  GRB,  estimated in the similar way  are also plotted in the figure.
The early post-maximum decline $\Delta m_{15}$ of quasi-bolometric light curve of SN 2014ad is slower 
($\Delta m_{15}$(Bol) = 0.80) than SNe 2009bb,  2012ap, and it is faster than 
SNe 1998bw,  2002ap. The late phase (between +30 d  and the last available data point) decline 
rate of SN 2014ad is 2.1 mag (100d)$^{-1}$, which is marginally faster  than SN 1998bw but slower than
SN 2009bb. The slower decline during the late phase indicates that the ejecta is still 
thick for $\gamma$-rays to escape. Further, inspite of the extremely  high expansion velocity during 
the early phase and relatively high velocity ($\sim$ 15000 km\,s$^{-1}$) during the late phase, the 
slowly declining bolometric light curve (indicative of efficient trapping of $\gamma$-rays) of SN 2014ad 
hints  towards a massive ejecta.   

An estimate of the contribution of UV  ($uvm2$  and $uvw2$) bands to bolometric luminosity is made. 
The first available  {\it Swift} UV data point corresponds to $\sim$ 3 d after maximum in the $B$ band.  
At this epoch  the UV  flux contributes $\sim$ 10\%  to the  bolometric flux. The UV contribution 
 decreases to $\sim$ 7\% at $\sim$ 15 days after $B$ maximum.   Thus, the contribution of UV band is small 
 even if $uvw2$/$uvm2$ brightness is overestimated due to contamination by the background light. There are only few broad-line  type Ic SNe for which UV observations are available,  again the temporal coverage for  most of 
them is sparse.  However, for SN 2006aj the contribution of   UV flux to bolometric flux  
is derived  by \citet{camp06} and  \citet{cano11}. They  showed that because of the shock breakout of the progenitor star, during the first few days  ($\sim$ 2 days) after the explosion, contribution of UV flux to the bolometric  flux was high, and thereafter decayed rapidly. For SN 2014ad, the estimated  UV contribution of $\sim$ 10\%  to bolometric flux at +3 d is consistent with the fraction estimated for SN 2006aj at 
similar epoch. In the quasi-bolometric curve presented in Fig. \ref{bol_lc1_ad}, no correction has been applied for the missing flux in the IR-bands.  It is shown that the  missing flux in NIR  bands can contribute $\sim$ 20 -- 25\% 
to the bolometric flux near maximum brightness, which increases to $\sim$ 40 -- 50\% one month later (\citealt{tomi06, vale08, cano11}). After including a contribution of 20\% for the missing band in NIR, the  peak bolometric magnitude  of SN 2014ad is -18.32 mag.

\begin{figure}
\centering
\resizebox{\hsize}{!}{\includegraphics{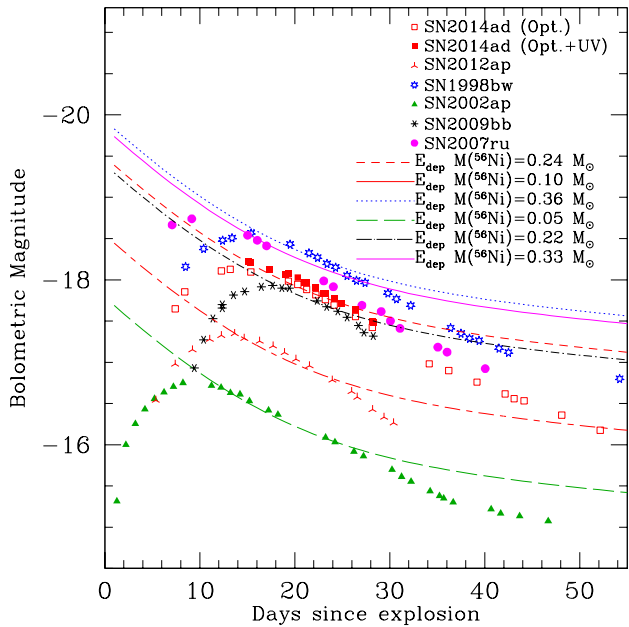}}
\caption[]{Quasi-bolometric light curve and of SN 2014ad and other broad-line type Ic SNe, the overplotted lines correspond to  rate of  energy production via  $^{56}$Ni $\rightarrow$  $^{56}$Co chain for different values of mass of $^{56}$Ni.}
\label{bol_lc1_ad}
\end{figure}

\section{Explosion parameters}

The bolometric light curve derived in Section \ref{sec:bol_curve} along with the information about the photospheric velocity is used to estimate explosion parameters {\it i.e.} mass of ejecta (M$_\text{ej}$),  kinetic energy  (E$_\text{k}$) and mass of $^{56}$Ni  synthesized in the explosion of  SN 2014ad. 
The explosion  parameters can be derived either by detailed hydroynamical light curve and spectral modelling of these events (\citealt{iwam98, mazz03,  mazz06,  vale08}), by fitting the analytical model of \citet{arne82} to the estimated  bolometric light curve of the supernova, or using the analytical formula for total rate of energy production via $^{56}$Ni $\rightarrow$  $^{56}$Co  \citep{nady94}. 

The  energy production  curves for  different values of mass of $^{56}$Ni has been estimated and are plotted in Fig. \ref{bol_lc1_ad} (thin line) alongwith the quasi-bolometric light curves. From the figure it is evident that the energy production rate for 0.24 M$_\odot$  of $^{56}$Ni  matches best to the quasi-bolometric light curve of SN 2014ad. For comparison, bolometric light curves and  the best matching  energy production  curves corresponding to different values of $^{56}$Ni  for some other well  studied broad-line Ic SNe are also shown in  Fig. \ref{bol_lc1_ad}, which are in good agreement with the values reported in the  literature.  

Alternatively, the  mass of $^{56}$Ni synthesized in the explosion can also be estimated using Arnett's formulation  (1982). This formulation is applicable to the SNe for which the light curves are powered purely by radioactive decay. The model assumes that during the photospheric phase  supernova ejecta expands homologously with spherical symmetry, $^{56}$Ni is located at the center without any mixing to outer layers, a single opacity over the duration of the explosion  and the ejecta is radiation-pressure dominated. 
The expression for luminosity as a function of time, modified by \citet{vale08}, to incorporate the energy produced by  decay of cobalt into iron, is used. The mass of  $^{56}$Ni  and $\tau_{m}$, the effective diffusion time that determines the overall width of the bolometric light curve, are the only free parameters in the expression for luminosity.    For a uniform density, the ejecta kinetic energy ${\rm E}_{\rm k}$ and the effective diffusion time scale $\tau_{m}$ (\citealt{arne82, arne96}) are  given by 
\begin{center}
\begin{equation}
{\rm E}_{\rm k}\approx \frac{3}{5} \frac{{{\rm M}_{\rm ej}v_{\rm ph}^{2}}}{2}
\end{equation} 
\end{center}
\noindent

\begin{center}
\begin{equation}
\label{equ:tau} 
\tau_{m} = \left(\frac{\kappa}{\beta c}\right)^{1/2} \left(\frac{{6 {\rm M}_{\rm ej}^{3}}}{{5 {\rm E}_{\rm k}}}\right)^{1/4}
\end{equation} 
\end{center}

where, $\kappa$ is the optical opacity, taken as 0.07 cm$^{2}$g$^{-1}$  \citep{chug00},   $\beta \approx 13.8$ is a  constant of integration \citep{arne82}  and  $c$ is the speed of light. 

The quasi-bolometric light curve of SN 2014ad is fit with the modified Arnett's formulation by varying mass of  $^{56}$Ni and  $\tau_{m}$. This formulation assumes that the material is in the photospheric phase, which holds good till $\sim$ 30 days, hence the bolometric light curve till  30 days was considered for fitting  Arnett's model.  The best  model fit  to the quasi-bolometric light curve is shown in Fig. \ref{arnett_fit}.  For the $UBVRI$ quasi-bolometric light curve, the best fit  values of $\tau_{m}$ and  mass of $^{56}$Ni are 11.6 $\pm$ 0.6  days and 0.22 M$_\odot$, respectively. The inclusion of {\it Swift} UV contribution to the quasi-bolometric light curve increases mass of  $^{56}$Ni to $\sim$ 0.24 M$_\odot$.  The mass of $^{56}$Ni derived using the energy deposition curve and that using Arnett's formulation  are in good agreement.

The photospheric velocity, at the time of bolometric maximum (13 d after explosion) is measured as  $v_{\rm ph}$  = 22300 $\pm$ 2100  km\,s$^{-1}$. The mass of 
ejecta estimated in this way is M$_\text{ej}$ = 3.3 $\pm$ 0.8 M$_\odot$ and the kinetic energy of explosion E$_\text{k}$ = (1 $\pm$ 0.3) $\times$ 10$^{52}$ ergs. The  errors in the explosion parameters are estimated by taking into account the errors in $\tau_{m}$ and measured photospheric velocity. The mass of $^{56}$Ni and the ejecta mass for SN 2014ad  matches well with the median values  for broad-line Ic supernova, however, the kinetic energy is much  higher than its  median values for broad-line Ic SNe (0.6$\times$10$^{52}$ ergs)  estimated  by \citet{cano13}.  

\begin{figure}
\centering
\resizebox{\hsize}{!}{\includegraphics{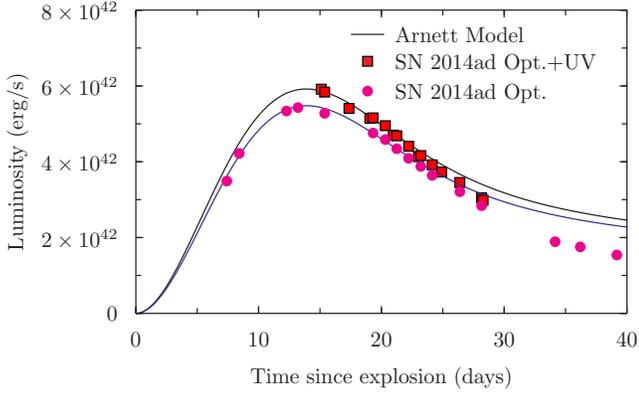}}
\caption[]{Arnett's analytical model fit to the quasi-bolometric light curve of SN 2014ad.} 
\label{arnett_fit}
\end{figure}

\citet{drou11}, have proposed an empirical relation 
between the mass of $^{56}$Ni synthesized in the explosion  and extinction corrected $R$ band absolute magnitude of the SNe,  
\begin{center}
\begin{equation}
 \log ({\rm M}_{\rm ^{56}Ni}/{\rm M}_{\odot}) = -0.41 \times M_{R} - 8.3
\end{equation}
\end{center}
where,  M$_{\rm ^{56}Ni}$ is the mass of $^{56}$Ni and $M_{R}$ is the extinction corrected $R$ band absolute magnitude of the supernova.  The extinction corrected $R$ band absolute magnitude of SN 2014ad is $M_{R} = -18.87 \pm$ 0.30 mag.  This leads to the mass of  $^{56}$Ni as 0.27 $\pm$ 0.06 M$_\odot$, which is consistent with other estimates of mass of $^{56}$Ni.  

\citet{cano13} suggests yet another method for estimating the explosion parameters of type Ibc SNe.
In this method, the stretch and luminosity factors in  $BVRI$ filter for a given supernova is determined relative to  $k$-corrected light curves of template supernova SN 1998bw. The bolometric light curve of SN 1998bw is then scaled with the average stretch and luminosity factors. The transformed bolometric light curve is fit with analytical model of \citet{arne82} and the mass of $^{56}$Ni, mass of the ejecta and kinetic energy of explosion are determined.  The advantage of using this method  lies on the fact that the explosion parameters can be determined for those objects also for which data is sparse or is not available in  all the optical bands. Using this method, \citet{cano13} has estimated  explosion parameters for a sample of GRB/XRF SNe, type Ib/c SNe and broad-line type Ic SNe.  
 
The explosion parameters from \citet{cano13} are plotted in Fig. \ref{mni_ekin}, \ref{mni_mej} and \ref{mej_ekin}. The parameters for SN 2014ad are also plotted in the same figures to check the position of SN 2014ad in these diagrams. It is pointed out by \citet{toy16}, that there was a typo in the expression of E$_\text{k}$ in \citet{arne82}, which was later corrected by \citet{arne96}. The  expression used for estimating  E$_\text{k}$ and and  M$_{\rm ej}$ in this work takes it into account.  As the   kinetic energy  E$_\text{k}$  was estimated by \citet{cano13} using E$_\text{k}= \frac{{{\rm M}_{\rm ej}v_{\rm ph}^{2}}}{2}$,  their  reported values of E$_\text{k}$  have been scaled by a factor of 3/5 to compare with this work. 

From Fig. \ref{mni_ekin} a correlation between the mass of $^{56}$Ni synthesized in the explosion and  kinetic energy of explosion E$_\text{k}$ is evident, objects with higher kinetic energy tend to produce more $^{56}$Ni in the explosion.  The type Ibc SNe are the explosions with lower kinetic energy and the mass of  $^{56}$Ni produced in them is also on the lower side, broad-line type Ic SNe exhibit higher kinetic energy and produce relatively larger $^{56}$Ni. As expected the GRB/XRF SNe  are associated with extremely high kinetic energy and produce large amount of $^{56}$Ni. In the E$_\text{k}$ - $^{56}$Ni plane, SN 2014ad appears closer to the GRB/XRF associated SNe than the broad-line Ic SNe. 
In Fig. \ref{mni_mej}  a correlation   between  mass of $^{56}$Ni and mass of ejecta is seen, though  it is not as strong as seen in the plot between mass of $^{56}$Ni and kinetic energy of explosion.  In this plot,    different subclass of SNe do not occupy distinct region and  there is a significant overlap. 

\begin{figure}
\centering
\resizebox{\hsize}{!}{\includegraphics{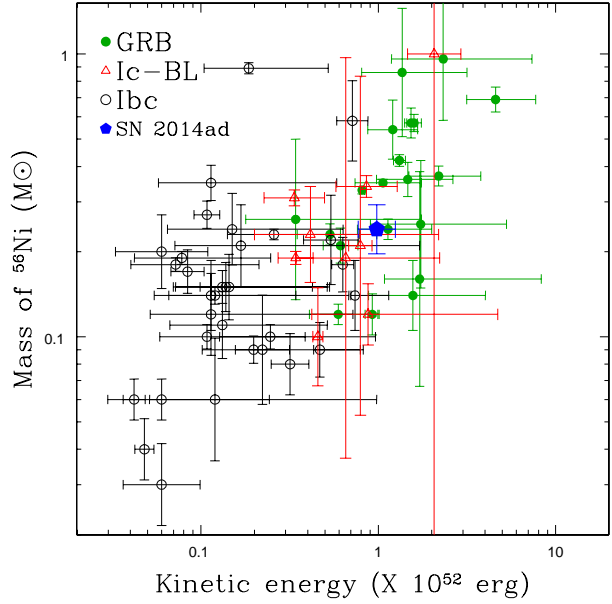}}
\caption[]{Mass of $^{56}$Ni synthesized in the explosion versus kinetic energy of explosion for GRB/XRF SNe, type Ib/c and broad-line type Ic SNe.} 
\label{mni_ekin}
\end{figure}

\begin{figure}
\centering
\resizebox{\hsize}{!}{\includegraphics{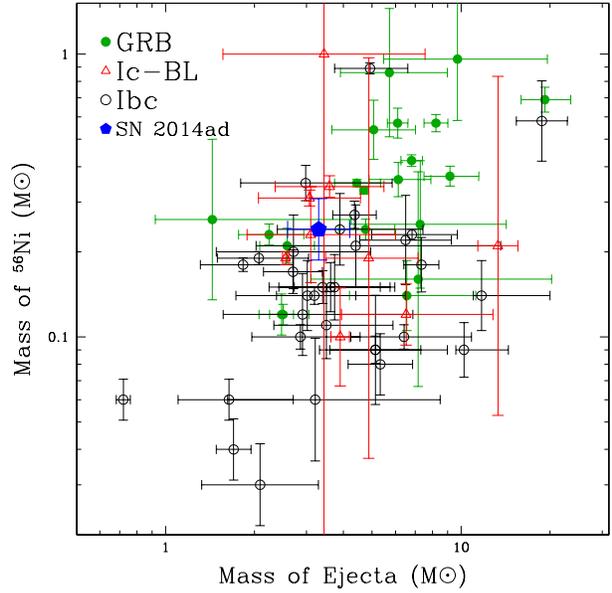}}
\caption[]{Mass of $^{56}$Ni synthesized in the explosion versus mass of ejecta for GRB/XRF SNe, type Ib/c and broad-line type Ic SNe.}
\label{mni_mej}
\end{figure}

In Fig. \ref{mej_ekin} two clear sequences are seen. There is a tight correlation between M$_\text{ej}$ and E$_\text{k}$ in case of type Ib/c SNe. Similarly, GRB/XRF SNe also show a tight correlation, but with a shift towards higher E$_\text{k}$/M$_\text{ej}$ ratio. Most of the broad-line Ic SNe are located in the region between the  Ibc SNe and GRB/XRF associated SNe. \citet{lyma16} have also shown similar plots in which they have included SNe IIb, Ib, Ic and broad-line Ic. The observed tight correlation between M$_\text{ej}$ and E$_\text{k}$ for SNe IIb, Ib and Ic was interpreted as the result of the similar photospheric velocity they exhibit. Though broad-line Ic SNe have comparable ejecta mass as those of SNe Ibc, because of their high expansion velocity they show high  E$_\text{k}$/M$_\text{ej}$ ratio. With relatively higher kinetic energy and lower ejecta mass,  SN 2014ad is located close to the strip occupied by GRB/XRF associated SNe. In  Fig. \ref{ekin_hist}, using the data from \citet{cano13},  distribution of Ib/c, broad-line Ic SNe and GRBs as a function of kinetic energy is plotted. In this distribution,  with kinetic energy  E$_\text{k}$ = (1 $\pm$ 0.3)$\times$10$^{52}$ ergs,  SN 2014ad lies close to the region occupied by GRBs. 
\begin{figure}
\centering
\resizebox{\hsize}{!}{\includegraphics{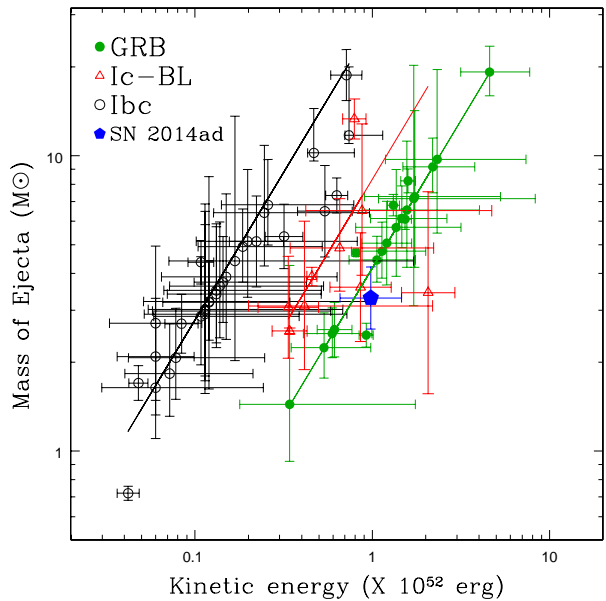}}
\caption[]{Mass of ejecta versus kinetic energy of explosion for GRBs, type Ib/c and broad-line type  Ic SNe. The solid lines correspond to the median value of E$_\text{k}$/M$_\text{ej}$ ratio for type Ib/c, Ic-BL SNe and GRBs.}
\label{mej_ekin}
\end{figure}

\begin{figure}
\centering
\resizebox{\hsize}{!}{\includegraphics{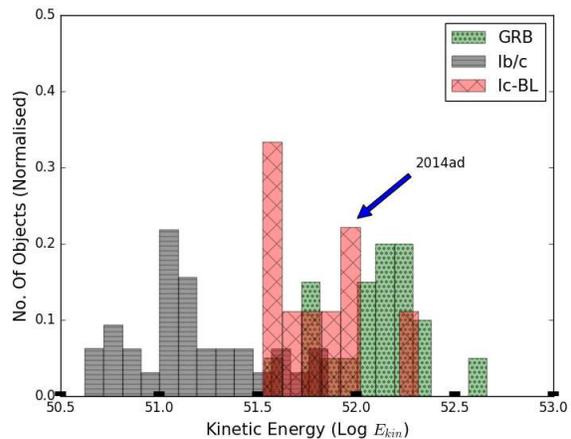}}
\caption[]{Distribution of  type Ib/c,  broad-line type  Ic SNe and GRBs as a function of kinetic energy.}
\label{ekin_hist}
\end{figure}

From the above discussion it is clear that  expansion velocity of the ejecta ($\sim$  22300 km\,s$^{-1}$ at bolometric peak) and the E$_\text{k}$/M$_\text{ej}$ ratio  ($\sim$ 0.30) of  SN 2014ad are  high.  It is larger than all other well studied broad-line Ic SNe (median value of E$_\text{k}$/M$_\text{ej}$ ratio $\sim$ 0.12).   
In fact, E$_\text{k}$/M$_\text{ej}$  ratio for SN 2014ad is even higher than GRB/XRF SNe (median value $\sim$ 0.24), except for SN 2010bh (GRB 100316D; \citealt{cano11}). We checked  for possible association of SN 2014ad  with GRB/XRF.  There is no report of  detection of any GRB from 19 February 2014, around the supernova location.   SN 2014ad was observed with the {\it Swift} X-ray Telescope (XRT; \citealt{burr05})  but there is no report of the detection of this supernova in X-rays too.  

There are two other broad-line type Ic supernovae, SN 2009bb \citep{pign11} and SN 2012ap \citep{mili15},  both bright in the radio  and mildly relativistic,  not associated with GRB/XRF,  did not show  X-ray emission at late times. In both cases, helium was detected in the early-time optical spectra with photospheric velocities  $>$  20000 km\,s$^{-1}$,  during  nebular phase the emission line flux ratio [O\,{\sc i}]/[Ca\,{\sc ii}] $<$ 1, and the metallicity of the supernova location is solar to super-solar. The explosion properties of both of these objects were shown to fall in between the normal Ibc events and SNe associated with GRBs.  They were proposed to be weak-engine driven explosions, where the engine activity  stops before being able to produce a successful jet breakout \citep{marg14}. As  the jet is not able to pierce through the stellar envelope only a small fraction of energy is dissipated at $\gamma$-ray frequencies, resulting in their non-detection in $\gamma$-ray region. 

The  factors which might be differentiating  between the broad-line Ic without GRB and GRB associated supernova are the different lifetime of the central engine and progenitor properties \citep{marg14}. The short lived engine may not be able to power the   jets sufficiently enough to pierce through the envelope of the progenitor to produce a GRB,  instead it can give rise to relativistic SNe \citep{lazz12}. It is shown that most of the observed GRBs are associated with the metal poor environment \citep{modj08}. However, discovery of GRBs  in high metallicity environment \citep{grah13} indicates that  though metallicity has some role,  it may not be the ultimate parameter deciding the final explosion outcome \citep{marg14}. The presence of helium is the other factor which affects breakout of jet.  Presence of helium in SN 2009bb and SN 2012ap is an indication that the helium layers in the progenitor of these events is not fully removed. \citet{marg14} have discussed the possibility that this helium layer may slow down the jet which subsequently  fails to outbreak through the progenitor and results in relativistic SNe.     

 Some of the properties of SN 2014ad,  namely, no association with GRB/XRF,    high ejecta velocity, weak/no X-ray emission, are common with SN 2009bb and SN 2012ap.  The nebular line flux ratio and metallicity in the supernova region of SN 2014ad  is higher than SN 2009bb and SN 2012ap. 
Inspite of  high kinetic energy and  high E$_\text{k}$/M$_\text{ej}$  ratio, no  GRB was detected with   SN 2014ad. One  reason for non detection of GRB  with SN 2014ad may be the jet pointing  away from the  line of sight of the observer.
 The other possibility is,  either  the central engine was weak and short lived or some mechanism, such as presence of thin helium layer, was operational which slowed down the jet and it failed to outbreak through the progenitor.

For the very energetic broad-line type Ic supernova SN 2010ah, the explosion parameters were estimated by \citet{mazz13} using  spectral and light curve modelling.  They  inferred SN 2010ah to be an explosion of a CO core of $\sim$ 5--6 M$_\odot$ and zero age main-sequence mass ($M_{MS}$)  of the progenitor star as 24--28 M$_\odot$.  Our estimates of the explosion parameters for SN 2014ad are similar to those of SN 2010ah indicating that the progenitor for SN 2014ad may also be a massive one. 
If we assume that a neutron star of $\sim$ 1.4--2.0 M$_\odot$ is formed, the CO core mass is $\sim$ 3.9--6.1 M$_\odot$.  This mass range corresponds to the main-sequence mass of $M_{MS} \sim$ 20--25 M$_\odot$, which is somewhat smaller than SN 2010ah. If we assume that a black hole more massive  than $\sim$ 3 M$_\odot$ is formed, the masses of the CO core $\geq$ 6 M$_\odot$,  then the $M_{MS} \geq$ 25 M$_\odot$.
Our inference of massive progenitor for SN 2014ad based on the nebular line ratio   [O\,{\sc i}]/[Ca\,{\sc ii}]  is consistent with this.

\section{Summary}
\label{sec:summary}
In this paper results based on low resolution optical spectroscopy,  optical imaging and  {\it Swift} UVOT imaging of broad-line type Ic supernova SN 2014ad have been presented. With an  absolute $V$ band magnitude of $-$18.86 $\pm$ 0.23 mag,   it is brighter than normal, and most of the broad-line  type Ic SNe,  but fainter than GRB associated SNe. The width of the spectral features and the expansion velocity of  the  ejecta is found to be very high. In the early phase the expansion velocity is as high as 0.1$c$, during the late phase  also the  expansion velocity is found to be higher than other objects of similar class except GRB associated SN 2010bh.  The explosion parameters determined by fitting Arnett's formulation  to the derived bolometric light curve shows it to be a highly energetic explosion, however,  no GRB/Xray flash was found to be associated with it.  The main sequence mass of the progenitor star is estimated to be $\geq$ 20 M$_\odot$.

\section*{Acknowledgement}
We thank  the referee for going through the manuscript carefully and providing constructive
comments, which improved the manuscript.  
All the observers of the 2-m HCT,  (operated by Indian Institute of Asrophysics), 
who kindly provided part of their observing time for observations
of the supernova, are thankfully acknowledged. We thank Takashi Hattori, Ji Hoon Kim and the
 staff at the Subaru Telescope for their excellent support of the observation under S15A-078 (PI: K. Maeda).
We thank Prof. David Branch for clarifying some of
the doubts related to spectrum synthesis code {\sc syn++}. 
We also thank Dr. Kuntal Mishra for providing us the code for fitting 
bolometric light curve.  This work has been supported in part by the DST-JSPS grant 
DST/INT/JSPS/P-211/2016, the Grant-in-Aid for Scientific Research of 
JSPS (15H02075, 16H02168,  26400222, 26800100) and MEXT (25103515, 15H00788) and by World Premier International
Research Center Initiative, MEXT, Japan. This work has made use of the public data in the
{\it Swift} data archive and the NASA/IPAC Extragalactic Database (NED) which is operated by Jet Propulsion 
Laboratory, California Institute of Technology, under contract with the National Aeronautics 
and Space Administration (NASA). 

\bibliographystyle{mn2e_warrick}
\bibliography{references.bib}

\label{lastpage}
\end{document}